\documentclass[a4paper]{article}
\usepackage{graphicx}
\usepackage{geometry}
\usepackage[version=4]{mhchem}
\usepackage{multirow}
\usepackage{leftindex}
\usepackage{subfigure}
\usepackage{amsfonts}
\usepackage{float}
\usepackage{rsc}
\usepackage{prettyref}
\newrefformat{fig}{Figure~\ref{#1}}
\newrefformat{table}{Table~\ref{#1}}
\newrefformat{eq}{Equation~(\ref{#1})}
\usepackage[colorlinks,linkcolor=black,citecolor=blue]{hyperref}
\newcommand{\tabincell}[2]{\begin{tabular}{@{}#1@{}}#2\end{tabular}}
\begin{document}
	\title{Comprehensive Molecular Representation from Equivariant Transformer}
	\author{Nianze Tao, Hiromi Morimoto and Stefano Leoni$^1$\footnote{email address: LeoniS@cardiff.ac.uk}}
	\date{%
    		$^1$Materials Discovery Group, School of Chemistry, Cardiff University, Cardiff, UK\\%
	}
	\maketitle
	
	\begin{abstract}
		The tradeoff between precision and performance in molecular simulations can nowadays be addressed by machine-learned force fields (MLFF), which combine \textit{ab initio} accuracy with force field numerical efficiency. Different from conventional force fields however, incorporating relevant electronic degrees of freedom into MLFFs becomes important. Here, we implement an equivariant transformer that embeds molecular net charge and spin state without additional neural network parameters. The model trained on a singlet/triplet non-correlated \ce{CH2} dataset can identify different spin states and shows state-of-the-art extrapolation capability. Therein, self-attention sensibly captures non-local effects, which, as we show, can be finely tuned over the network hyper-parameters. We indeed found that Softmax activation functions utilised in the self-attention mechanism of graph networks outperformed ReLU-like functions in prediction accuracy. Increasing the attention temperature from $\tau = \sqrt{d}$ to $\sqrt{2d}$ further improved the extrapolation capability, indicating a weighty role of nonlocality. Additionally, a weight initialisation method was purposed that sensibly accelerated the training process.
	\end{abstract}
	
	\section{Introduction}
	The need to expand the scope of molecular dynamics simulations has accelerated the resolution of the historical tradeoff between accuracy and efficiency in calculating interatomic forces~\cite{Unke_chemrev_2021}. Ab initio approaches encode the relevant physical picture, however solving the Schr\"odinger equation, albeit approximately, entails severely constraining system size and time scope of any simulation. Classical force fields on the contrary focus on the dependency of the potential energy on atomic coordinates via an analytical representation. Early applications of MLFFs to molecular simulations of bulk silicon~\cite{Behler_Si_PRL} expressed total energies as sum of atomic energies depending on local atomic coordination, encoded by symmetry functions, fed into the neural network. 
	Graph-based neural networks have recently achieved significant success in learning and predicting molecular energies and/or interatomic forces. Different architectures including invariant networks\cite{schutt2017schnet,unke2019physnet,gasteiger2020directional,unke2021spookynet,fan2021neuroevolution}, covariant networks\cite{anderson2019cormorant}, and equivariant networks\cite{schutt2021equivariant,tholke2022torchmd,satorras2021n,hutchinson2021lietransformer,batatia2022mace,batzner20223,haghighatlari2022newtonnet,fuchs2020se,brandstetter2021geometric,thomas2018tensor,batatia2022design}, which predict molecular energy and forces from raw atomic numbers and positions (Cartesian coordinates), showed higher computing efficiency than \textit{ab initio} quantum chemistry methods, without accuracy deterioration. Among these architectures, equivariant message passing neural networks (MPNN) have been shown to reliably predict tensor properties (e.g., dipole moment) with higher accuracy\cite{schutt2021equivariant}, which prompted further efforts to develop different models based on various equivariant groups\cite{schutt2021equivariant,tholke2022torchmd,hutchinson2021lietransformer,batatia2022mace}. It was shown \cite{schutt2021equivariant,tholke2022torchmd} that the equivariant characteristic even benefited the prediction of invariant properties (e.g., total energy of a molecule). Recent works also successfully combined equivariant message passing with transformer\cite{tholke2022torchmd,hutchinson2021lietransformer} achieving significant advances. Therein, P. Th{\"o}lke \textit{et al}\cite{tholke2022torchmd} pointed out the key role of activation functions in self-attention mechanism, preferring ReLU-like functions for improved accuracy. 
	
	Attention temperature\cite{zhang2021attention} in the self-attention mechanism was shown to play an important role in performance generalisation  in NLP models\cite{zhang2021attention}. For chemical system in particular, self-attention is believed to play a major role in learning non-local effects implied by electronic degrees of freedom, a feature that may improve transferability\cite{unke2021spookynet}.
	
	As mentioned by O.T. Unke \textit{et al}\cite{unke2021spookynet}, the atomic numbers (Z) and coordinates (R) are not a complete representation of a molecule. A comprehensive representation, e.g., a wavefunction, besides Z and R, must  include the molecule net charge (Q) and total spin (S) as well. Along this line, here we present Comprehensive Molecular Representation from Equivariant Transformer (CMRET), which explicitly embeds Z, R, Q and S into the model. Different from SpookyNet\cite{unke2021spookynet} that embeds Q and S with Z, employing an attention-like strategy that requires adding more parameters into the neural network, our purposed method is trainable parameter-free and can be easily applied to other equivariant models (e.g., PAINN\cite{schutt2021equivariant}) using vector features, like demonstrated in the following. 
	
	\section{Methods}
	\subsection{Architecture of Equivariant Transformer}
	
\begin{figure}[ht]
		\centering
		\includegraphics[width=0.95\linewidth]{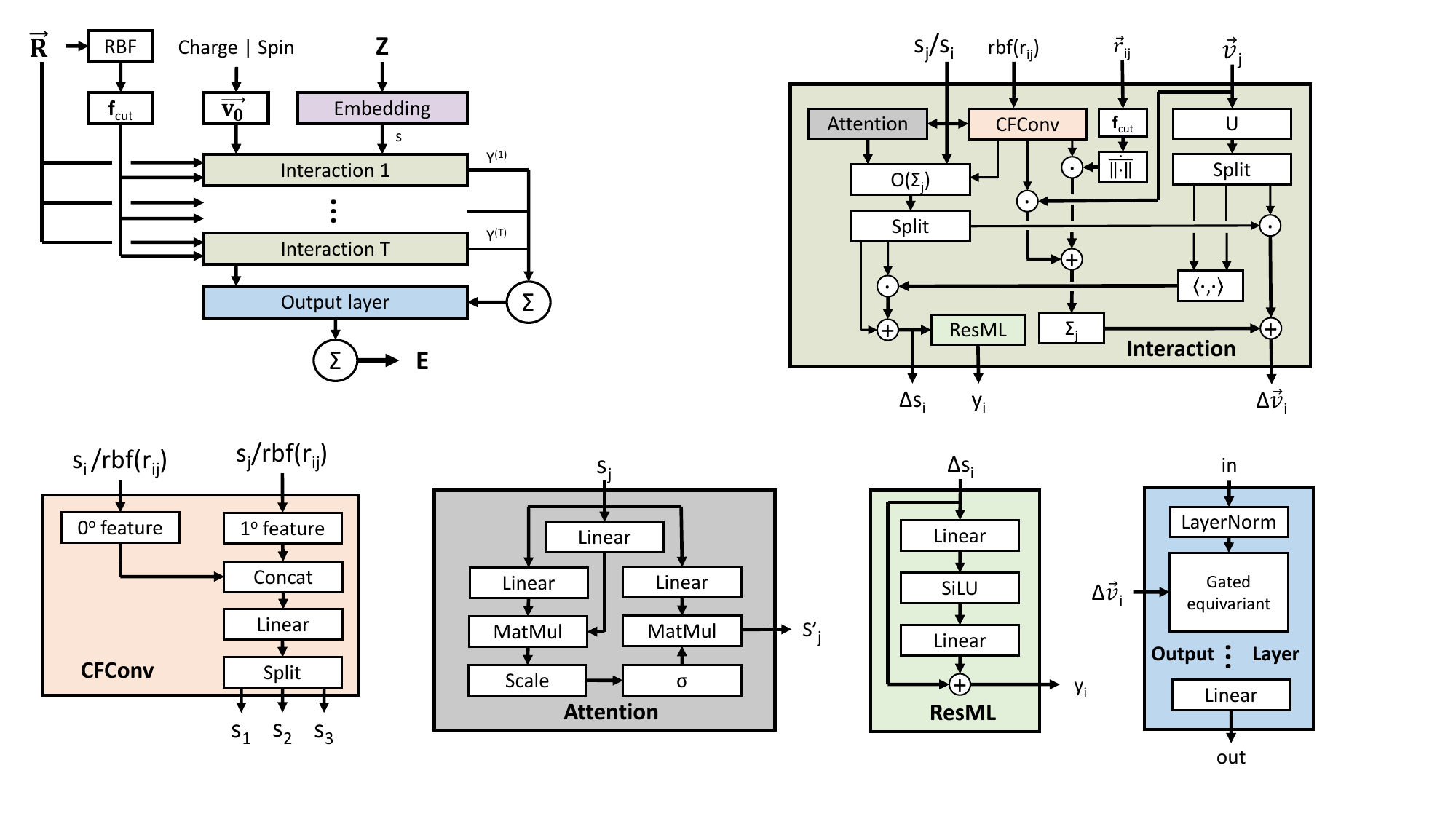}
		\caption{Colour-coded overview of the structure of CMRET and its individual components. Thin lines represent split scalar or vector features. `Split' method cleaves the input feature into 3 equal-length outputs. `Scale' method scales the input to its 1/$\tau$ ($\tau$ $>$ 1). $\sigma$ represents any activation function. The structure of Gated equivariant block introduced by K.F. Sch{\"u}tt \textit{et al}\cite{schutt2021equivariant} is not shown here.}
		\label{fig:nn}
	\end{figure}

	In this section, we detail key CMRET componets, i.e., embedding, radial basis function (RBF), interaction block, and output layer. The CMRET net takes atomic numbers and Cartesian coordinates as input. To achieve a wavefunction representation, the total charge and total spin are supplemented as vector feature prior to embedding. The network overview scheme is shown in \prettyref{fig:nn}. 


	\subsubsection{Embedding}
	The embedding layer first maps the atomic number $z$ to the ground state electron configuration (from 1s orbital to 7f orbital) $x^{embedded}$ of the element, which then is passed through a linear layer, i.e.,
	\begin{equation}
		s_{i}^{0} = embed(z_{i}) = W^{embed}x_{i}^{embedded} + b^{embed}.
	\end{equation}
	$s \in \mathbb{R}^{\textrm N_{atom}\times F}$ is the scalar feature of the included atoms, in which F is the number of atomic features (defaulting to 128).


	\subsubsection{Radial Basis Functions (RBF)}

	Two different radial basis functions are implemented in CMRET: a Bessel type\cite{schutt2021equivariant,gasteiger2020directional} RBF and a Gaussian type\cite{tholke2022torchmd,unke2019physnet} RBF. The Bessel type RBF is defined as
	\begin{equation}
		RBF(d_{ij}) = \frac{1}{d_{ij}}sin(\frac{n\pi}{d_{cut}}d_{ij}),
	\end{equation}
	where $d_{cut}$ is the cut-off radius, $d_{ij} = \|\vec{r}_{ij}\| = \|\vec{r}_{i}-\vec{r}_{j}\|$ is the pair-wise distance between atom $i$ and atom $j$, and $n$ is an integer number $n \in [1, N_{basis}]$. The Gaussian type RBF is defined as
	\begin{equation}
		RBF(d_{ij}) = e^{-\beta_{n}(e^{-d_{ij}} - \mu_{n})^{2}},
	\end{equation}
	where $\beta_{n}$ and $\mu_{n}$ are trainable parameters: all $\beta$ values are initialised as $(\frac{N_{basis}}{2})^{2}(1 - e^{-d_{cut}})^{-2}$ and $\mu_{n} \in \{\mu_{1}, \mu_{2}, \mu_{3}, ..., \mu_{N_{basis}}\}$ in which for $\forall i,j$ $\|\mu_{i}-\mu_{j}\| = (1 - e^{-d_{cut}})/N_{basis}$ as initial values. The value of $N_{basis}$ defaults to 50 in CMRET.
	
	A cosine cut-off function, i.e.,
	\begin{equation}
		\phi(d_{ij}) = \left\{\begin{array}{llc}
			{\frac{1}{2}\left(cos\left(\frac{\pi}{d_{cut}}d_{ij}\right) + 1\right)} & {,} & {d_{ij} \leq d_{cut}}\\
			{0} & {,} & {d_{ij} > d_{cut}}\end{array}\right.
		\label{eq:cutoff}
	\end{equation}
	is operated onto the RBF to create the edge feature as
	\begin{equation}
		e_{ij} = \phi\circ RBF(d_{ij}).
	\end{equation}
	The attention mechanism, discussed below, allows for non-linear mixing of the basis functions relative to the basis vectors of the coordinate frames, allowing to naturally incorporating an angular dependency.


	\subsubsection{Attention}
	The attention layer follows the design of scaled dot-product self-attention\cite{vaswani2017attention} mechanism, i.e.,
	\begin{equation}
		s'_{i} = attention(s_{i}) = \sum_{j}^{all}\alpha_{ij}(W^{V}s_{j} + b^{V})
	\end{equation}
	\begin{equation}
		\alpha_{ij} \in A = \sigma[(W^{Q}s + b^{Q})(W^{K}s + b^{K})^{\rm T}/\tau],
		\label{eq:alpha}
	\end{equation}
	where $\sigma$ is the attention activation function and $\tau$ is the attention temperature ($\tau > 0$). A multi-head attention mechanism is also implemented in our model (4 heads as default).
	\par This module is parallel to the scalar-level message-passing block (i.e., CFConv block), which is expected to capture the long-range interactions outside the cut-off range defined in \prettyref{eq:cutoff}. 


	\subsubsection{Modified Continuous-Filter Convolution (CFConv)}
	In the normal continuous-filter convolution introduced in SchNet\cite{schutt2017schnet}, the features of first order neighbourhood nodes and edges (i.e., entities within the range of $d_{cut}$) are aggregated together. Inspired by MixHop\cite{abu2019mixhop} convolution, we add the zero order (self) features:
	\begin{equation}
		\begin{array}{lcl}
			{\leftindex[I]^{l}s_{ij}^{nbh}} & = & {\leftindex[I]^{l-1}{\hat{s}_{j}}\odot SiLU(\leftindex[I]^{l}W^{e_{1}}e_{ij} + \leftindex[I]^{l}b^{e_{1}})\odot\phi_{ij}}\\
			{\leftindex[I]^{l}s_{ij}^{self}} & = & {\leftindex[I]^{l-1}{\hat{s}_{i}}\odot SiLU(\leftindex[I]^{l}W^{e_{2}}e_{ij} + \leftindex[I]^{l}b^{e_{2}})\odot\phi_{ij}}\\
			{\leftindex[I]^{l}s_{ij}} & = & {\leftindex[I]^{l}W^{o}\left( \leftindex[I]^{l}s_{ij}^{self}\|\leftindex[I]^{l}s_{ij}^{nbh} \right)+\leftindex[I]^{l}b^{o}},
		\end{array}
	\end{equation}
	where $\odot$ represents the element-wise product, $\|$ is the concatenation operator over feature dimension, and $\leftindex[I]^{l-1}{\hat{s}_{i}} = SiLU(\leftindex[I]^{l}W^{\Phi}\leftindex[I]^{l-1}s_{i} + \leftindex[I]^{l}b^{\Phi})$ ($W^{e_{1}},W^{e_{2}} \in \mathbb{R}^{N_{basis}\times F}$, $W^{\Phi} \in \mathbb{R}^{F\times F}$ and $W^{o} \in \mathbb{R}^{2F\times 3F}$). A cutoff function $\phi$ is used to smooth the potential. Scalar features $s_{1}$, $s_{2}$ and $s_{3}$ are generated by equally splitting $s$.


	\subsubsection{Interaction}
	The Interaction block utilises an equivariant topology similar to Torchmd-NET\cite{tholke2022torchmd} when updating the scalar feature and vector feature. However, instead of filling the initial vector feature $\vec{\rm V}_{0} \in \mathbb{R}^{\rm N_{atom}\times 3\times F}$ with zero matrix elements, we embed the information of molecular net charge and spin into the feature:
	\begin{equation}
		\vec{\rm V}_{0} = \vec{0} - \frac{charge - spin}{\sum_{i}^{all}z_{i}}\vec{1} = [\vec{v}_{1}\; \vec{v}_{2}\; ...\; \vec{v}_{\rm N_{atom}}]^{\rm T},
	\end{equation}
	where $\vec{1}$ is an all-one matrix.
	
	The atomic interaction is separated into local interaction (modified continuous-filter convolution) and non-local interaction (self-attention). The visualised scheme is shown in \prettyref{fig:nn}. The updated scalar feature $\Delta s_{i}$ is
	\begin{equation}
		\Delta s_{i} = \left[O^{1}\left(\sum_{j}^{\forall j, d_{ij}\leq d_{cut}}s_{1}^{ij} + s_{i} + s'_{i}\right) + b^{1}\right] + \left[O^{2}\left(\sum_{j}^{\forall j, d_{ij}\leq d_{cut}}s_{1}^{ij} + s_{i} + s'_{i}\right) + b^{2}\right]\odot \left\langle U^{1}\vec{v}_{i}, U^{2}\vec{v}_{i}\right\rangle,
	\end{equation}
	where $\langle a, b\rangle$ is the inner product between $a$ and $b$. Then $y_{i} = ResML(\Delta s_{i})$.
	
	The updated vector feature $\Delta\vec{v}_{i}$ is 
	\begin{equation}
		\Delta\vec{v}_{i} = \sum_{j}^{\forall j, d_{ij}\leq d_{cut}}(s_{2}^{ij}\odot\vec{v}_{i} + s_{3}^{ij}\odot\frac{\vec{r}_{ij}}{\|\vec{r}_{ij}\|}) + \left[O^{3}\left(\sum_{j}^{\forall j, d_{ij}\leq d_{cut}}s_{1}^{ij} + s_{i} + s'_{i}\right) + b^{3}\right]\odot U^{3}\vec{v}_{i},
	\end{equation}
	where $O^{1},O^{2},O^{3},U^{1},U^{2},U^{3}\in\mathbb{R}^{F\times F}$ are the weights of linear layers.
	
	In our model, we used 6 Interaction layers (T = 6), unless otherwise specified.

	
	\subsubsection{Output Layer}
	The output layer consists of three parts: a layer normalisation\cite{ba2016layer} block, a series of Gated Equivariant\cite{schutt2021equivariant} blocks (default to use 2 blocks) and two fully connected layers mapping atomic feature number F $\rightarrow$ 1, i.e., $s^{out}_{i}=W^{s}s_{i}$ and $\vec{v}^{out}_{i}=W^{v}\vec{v}_{i}$. For the pure scalar feature (e.g., total energy, $\epsilon_{\rm HOMO}$), the output is simply $E=\sum_{i}^{all}s^{out}_{i}$ ($\vec{F}=-\nabla E$ if the forces are required). The electronic spatial extent is
	\begin{equation}
		\langle R^{2} \rangle=\sum_{i}^{all}s^{out}_{i}\|(\vec{r_{i}}-\vec{r_{0}})\|
	\end{equation}
	where $\vec{r}_{0}=\frac{1}{\sum_{i}^{all}z_{i}}\sum_{i}^{all}z_{i}\vec{r}_{i}$ is the charge centre. The computations of tensor features employ the vector output $\vec{v}^{out}$ from the model. For the calculation of dipole moment, the output is defined as
	\begin{equation}
		\mu=\|\vec{\mu}\|=\left\| \sum_{i}^{all}\vec{v}^{out}_{i}+s^{out}_{i}(\vec{r_{i}}-\vec{r_{0}}) \right\|.
	\end{equation}
	The isotropic polarisability is computed as
	\begin{equation}
		\alpha = \left\| \sum_{i}^{all}s^{out}_{i}I_{3}+\vec{v}^{out}_{i}\otimes (\vec{r_{i}}-\vec{r_{0}})+(\vec{r_{i}}-\vec{r_{0}})\otimes\vec{v}^{out}_{i} \right\|,
	\end{equation}
	where $\otimes$ stands for the outer product, and $I_{3}$ is the $3\times 3$ identity matrix.
	\subsection{Data Preparation}
	A singlet/triplet \ce{CH2} dataset containing 2,000 data points was generated based on DFT calculations at B3LYP\cite{becke1993becke,PhysRevB.37.785}/cc-pVDZ level of accuracy, in order to test the capability of CMRET to learn from electronic degrees of freedom. By sampling the 2 C-H bond lengths from 0.95 to 1.20 $\textrm \AA$ and the H-C-H angle from 90\textdegree\;to 140\textdegree, 1,000 uncorrelated geometries were generated, from which energies and forces where calculated.
	
	1,500 randomly selected data points (750 singlet + 750 triplet) constituted the training set. The remaining 500 geometries were used to test the performance of CMRET.


	\subsection{Training}
	The model can be trained on both scalar and vector features, if vector features exist in the dataset. The training loss is defined as
	\begin{equation}
		loss = \left\{\begin{array}{lcr}{0.2\cdot MSE(S, \hat{S}) + 0.8\cdot MSE(\vec{V}, \hat{\vec{V}})} & \mbox{if} & {\exists\vec{V}}\\ MSE(S, \hat{S}) & \mbox{if} & {\lnot\exists\vec{V}} \end{array}\right.,
	\end{equation}
	where $\hat{S}$ and $\hat{\vec{V}}$ are labelled scalar and labelled vector properties, respectively. The optimiser is Adam\cite{kingma2014adam} with $\beta_{1} = 0.9$, $\beta_{2} = 0.999$ and $\epsilon = 10^{-8}$. We found that AMSGrad\cite{j.2018on} strategy had negative effect on training results. A cyclic learning rate scheduler\cite{smith2017cyclical} (cycle\_momentum=False) is utilised to guide the learning rate for 2 cycles: during the first cycle the learning rate range was [$10^{-6}$, $10^{-4}$], which in the second cycle was reduced to [$10^{-8}$, $10^{-6}$]. The batch-size was 10 for singlet/triplet \ce{CH2} dataset, 5 for MD17\cite{doi:10.1126/sciadv.1603015} dataset, 20 for DES370K\cite{donchev2021quantum} dataset and 15 for ISO17\cite{schutt2017schnet} and QM9\cite{ramakrishnan2014quantum} dataset.


		\subsection{Weight Initialisation}
	A weight initialisation strategy was developed to stabilise and fasten the training process. We initialised the filter network in CFConv ($W^{e_{1}}$ and $W^{e_{2}}$) with an uniform distribution, in which the values were chosen between $-\sqrt{6/N_{basis}}$ and $\sqrt{6/N_{basis}}$. All the remaining linear layers before a SiLU activation function were initialised with a normal distribution with $\mu = 0$ and $\sigma = \sqrt{3/2F}$.

\section{Experiments and Results}

Learning molecular data entails encoding equilibrium geometries as well as not-equilibrium geometries, which can be systematically obtained from molecular dynamics simulation protocols. The reliable prediction of energies and forces sensibly depends on the sensitivity of the learning framework to changes in molecular internal coordinates. To disentangle the contribution of attention layer from output layer, we first performed an ablation of those component individually from the CMRET network, followed by a focused investigation on the role of activation function ($\sigma$ in \prettyref{fig:nn}) and activation temperature for singlet/triplet carbene. Further, we trained and tested CMRET on static and dynamic data, including the calculation of static properties, and compared our results to state-of-the-art networks. Therein, we focused on extrapolation capabilities on unseen data (unknown molecules and conformations), either as single geometries, dimers or as trajectories. 

\subsection{Ablation Tests}
\begin{figure}[ht]
		\subfigure[Fully structured CMRET (2 gated equivariant blocks in the output layer) \textit{vs.} DFT.]
		{\begin{minipage}[c][1\width]{0.31\textwidth}
				\centering
				\includegraphics[width=1\textwidth]{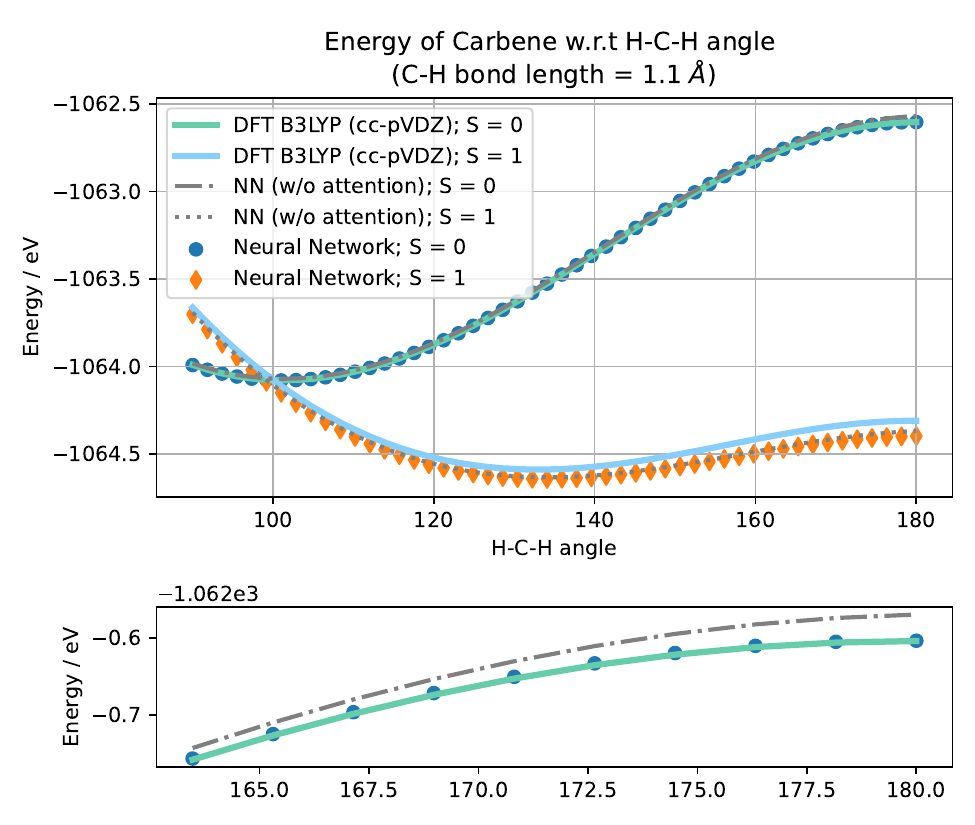}
				\label{fig:nn_t5}
		\end{minipage}}		
		\hfill
		\subfigure[CMRET with simplified output layer (1 gated equivariant block) \textit{vs.} DFT.]
		{\begin{minipage}[c][1\width]{0.31\textwidth}
				\centering
				\includegraphics[width=1\textwidth]{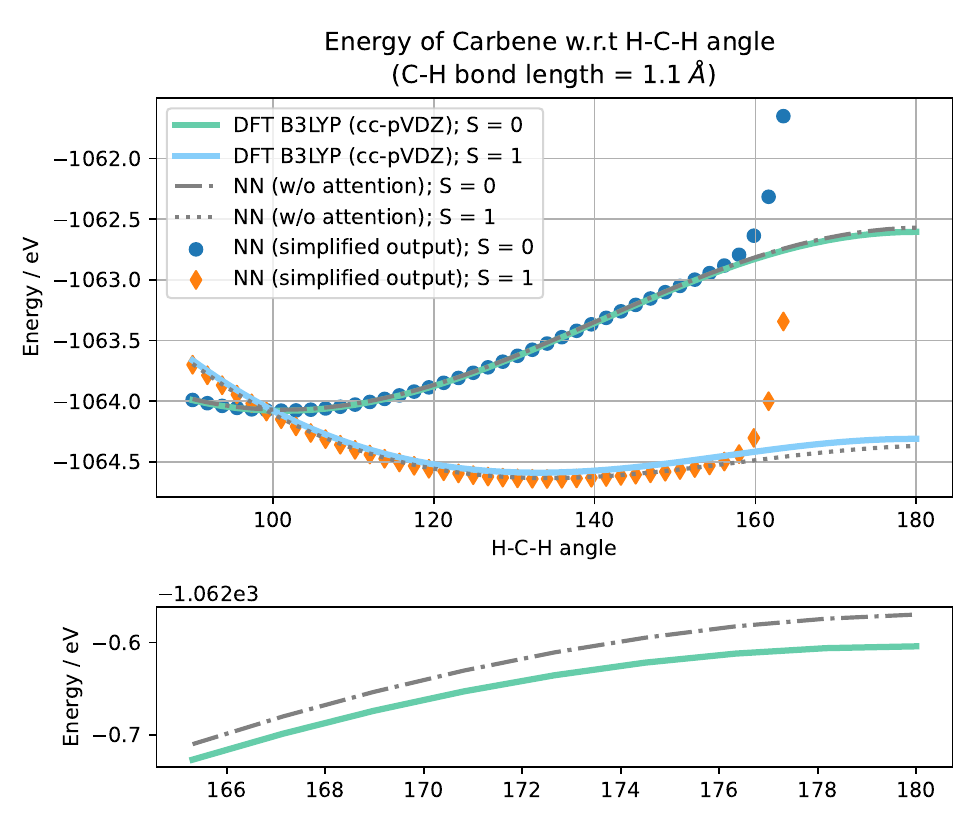}
				\label{fig:nn_t5_1}
		\end{minipage}}
		\hfill
		\subfigure[CMRET with simplified output (1 gated equivariant block) and interaction layers (no ResML structures) \textit{vs.} DFT.]
		{\begin{minipage}[c][1\width]{0.31\textwidth}
				\centering
				\includegraphics[width=1\textwidth]{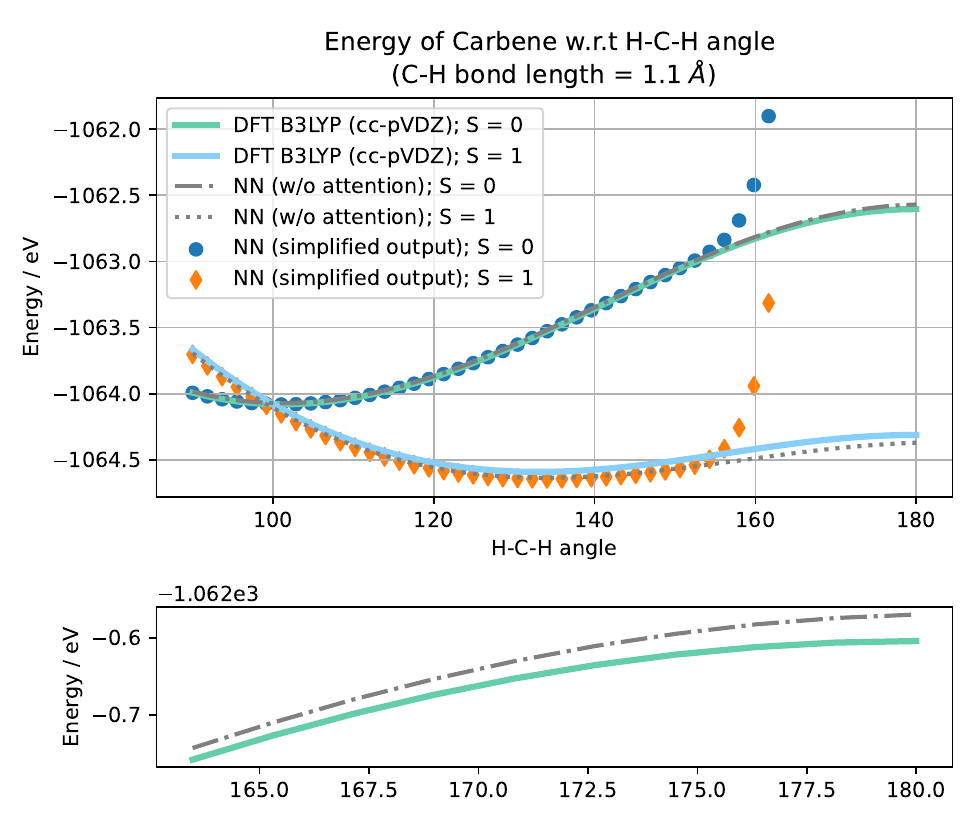}
				\label{fig:nn_t5_2}
		\end{minipage}}
\caption{Total energies of singlet/triplet \ce{CH2} calculated by CMRET with different output layer structures and DFT method (B3LYP/cc-pVDZ) scanning through the H-C-H angle from 90 to 180\textdegree\;with C-H bond length fixed at 1.1 $\textrm \AA$. Dashed grey lines represent the energies calculated by the neural network with fully structured output layer, but without applying self-attention networks.}
	\end{figure}
	
	We evaluated the effects of attention layer and the structure of output layer on the extrapolation capability of the model on the singlet/triplet carbene dataset. Eliminating the attention layer had a negative effect on the prediction accuracy when the H-C-H was larger than 165\textdegree\;(\prettyref{fig:nn_t5}), but the prediction values of the model still followed the trend of DFT. However, reducing the number of Gated Equivariant blocks in the output layer significantly weakened the extrapolation capability; removing ResML blocks made the performance worse (\prettyref{fig:nn_t5_1} \& \prettyref{fig:nn_t5_2}). These observations leads to two insights: (1) the self-attention mechanism is a useful correction to the message-passing processes and (2) the severe influence of parameters outside Interaction blocks (message-passing + self-attention) on the extrapolative behaviours is a hint that the algorithm in the Interaction layer is sufficient to encode the information of a molecular graph into atom-wise information (i.e., atomic latent vectors), to which enough interpretation power of the output layer (i.e., the number of parameters in a machine learning point of view) is required to extract the molecular property.  
	\subsection{The Attention Activation Function and Attention Temperature}
	\begin{figure}[ht]
		\centering
		\includegraphics[width=1.0\linewidth]{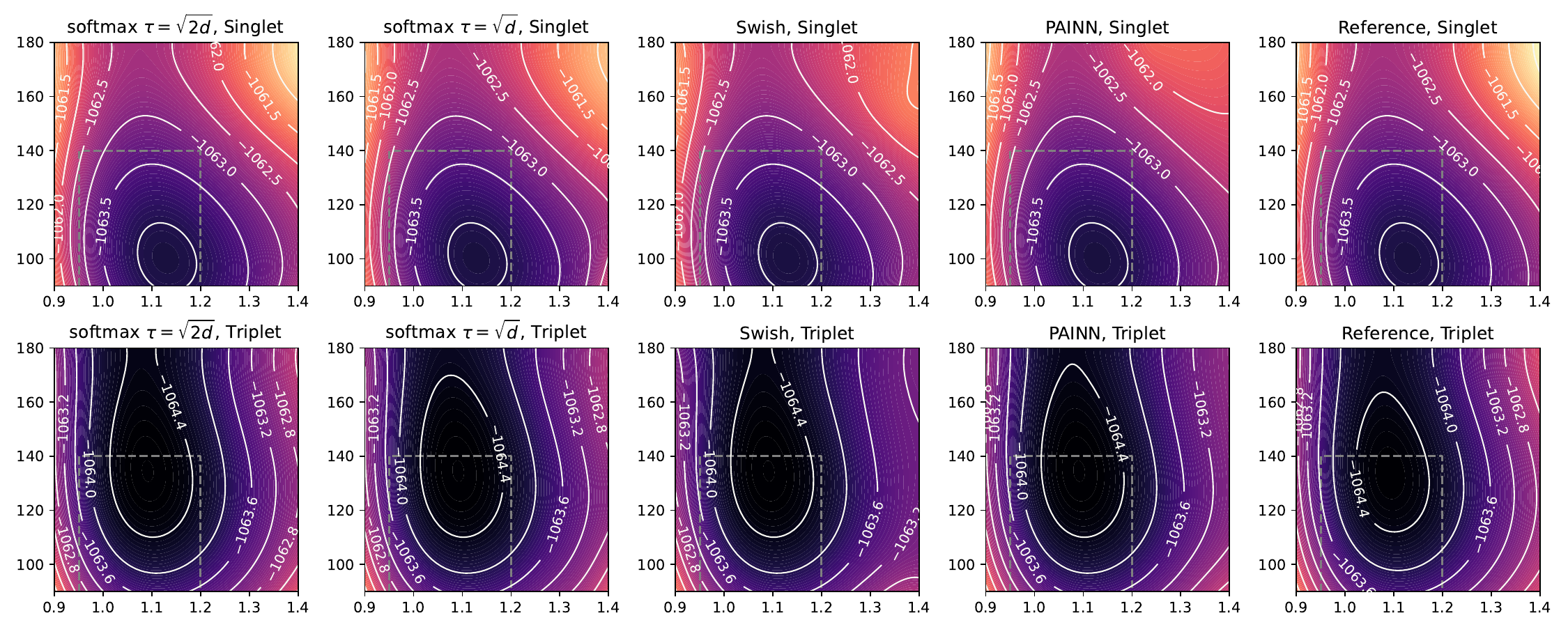}
		\caption{Potential surface of singlet/triplet \ce{CH2} calculated by CMRET (attention $\sigma$ = softmax with $\tau$ = $\sqrt{2d}$ and $\tau$ = $\sqrt{d}$, and $\sigma$ = Swish), modified PAINN, and DFT (B3LYP/cc-pVDZ) from left to right. X-axis is the C-H bond length from 0.9 to 1.4 $\rm \AA$ (the two C-H bonds have equal length in this case). Y-axis is the H-C-H angle varied from 90\textdegree\;to 180\textdegree. A brighter colour represents higher energy. Geometries in the training and testing datasets are included in the square delimited by dashed grey lines. All the neural network models were trained to converge.}
		\label{fig:ch2_2d}
	\end{figure}
	We tested the performance of our model on singlet/triplet \ce{CH2} set on applying different activation functions in the attention layer, including Softmax (softmax function operating on the last dimension), Softmax2d (softmax function operating on the last two dimensions), and a range of ReLU-like smooth and continuous functions: Swish\cite{ramachandran2017searching}, Softplus, GELU\cite{hendrycks2016gaussian} and Mish\cite{misra2019mish}. The comparison can be found in \prettyref{table:eps1} \& \prettyref{table:eps2}. We employed single-head attention, T = 5, and $N_{basis}=20$ in this test.
	
	\prettyref{table:eps2} also shows the effect of attention temperatures of $\tau = \sqrt{d}$, $\sqrt{1.5d}$ and $\sqrt{2d}$, where $d$ is the feature number in the attention layer. The out-of-domain test (results showed in \prettyref{fig:ch2_2d}) gave clearer information about the benefit of increasing the attention temperature. By scanning the potential surface of carbene, in which a large fraction of conformations were unseen during training, we found that the combination of $\sigma$ = Softmax, $\tau = \sqrt{2d}$ in the attention layer provided the best extrapolation capability (here, to enable PAINN to learn molecular spin states, we applied the same charge/spin embedding strategy of CMRET; the original nuclear embedding and output layer of PAINN were also replaced by the same method used in our model), as the structure of the potential energy surface (left in \prettyref{fig:ch2_2d}) was closer to ground-truth energy (right in \prettyref{fig:ch2_2d}) than the results obtained with other settings.
	
	From this test we concluded that Softmax function in the attention mechanism improves prediction accuracy, compared to ReLU-like functions. Increasing the attention temperature to $\tau = \sqrt{2d}$ promotes the generalisation capability of the trained model. We observed that even with more trainable parameters, the multi-head attention models with $\tau = \sqrt{d}$ did not outperform the single-head models. 
	\begin{table}[H]
		\centering
		\begin{tabular}{|c|cccccc|}
			\hline
			{} & {Swish} & {Softmax} & {Softplus} & {GELU} & {Mish} & \multirow{2}*{\tabincell{c}{Multi-head attention\\\scriptsize(2 heads)}}\\
			{} & {} & {} & {} & {} & {} & {}\\
			\hline
			{energy} & {5.795} & {\bf 0.458} & {0.624} & {2.368} & {1.914} & {3.569}\\
			{forces} & {0.292} & {\bf 0.212} & {0.215} & {0.244} & {0.264} & {0.295}\\
			\hline
		\end{tabular}
		\caption{Testing results on singlet/triplet \ce{CH2} data (metric: MAE) with different attention activation functions. The best result is labelled in bold. The models were trained for 7,000 epochs. The RBF type was Bessel. The unit of energy is meV and the unit of forces is meV/$\rm \AA$. The multi-head attention\cite{vaswani2017attention} mechanism was directly imported from the implementation in PyTorch.}
		\label{table:eps1}
	\end{table}
	\begin{table}[H]
		\centering
		\begin{tabular}{|l|ccccc|c|}
			\hline
			{} & {} & {\small Swish} & {\small Softmax} & {\small Softmax2d} & \multirow{2}*{\tabincell{c}{\small Multi-head attention\\\scriptsize(4 heads)}} & \multirow{2}*{\tabincell{c}{\small W/o\\\small attention}}\\
			{} & {} & {} & {} & {} & {} & {}\\
			\hline
			{\small\# parameter} & {} & {1.48M} & {1.48M} & {1.48M} & {1.81M} & {1.23M}\\
			\multirow{2}*{\tabincell{c}{$\tau = \sqrt{d}$\\}} & {\small energy} & {0.041} & {\bf 0.039} & {0.158} & {0.055} & {0.101}\\
			{} & {\small forces} & {0.156} & {0.150} & {0.151} & {0.151} & {0.161}\\
			\multirow{2}*{\tabincell{c}{$\tau = \sqrt{1.5d}$\\}} & {\small energy} & {--} & {0.351} & {--} & {--} & {}\\
			{} & {\small forces} & {--} & {0.148} & {--} & {--} & {}\\
			\multirow{2}*{\tabincell{c}{$\tau = \sqrt{2d}$\\}} & {\small energy} & {0.048} & {0.047} & {--} & {--} & {}\\
			{} & {\small forces} & {0.150} & {\bf 0.147} & {--} & {--} & {}\\
			\hline
		\end{tabular}
		\caption{Testing results on singlet/triplet \ce{CH2} data (metric: MAE) with different attention activation functions and different attention temperatures. The best result is labelled in bold. The models were trained for 20k epochs. RBF type was Gaussian. The unit of energy is meV and the unit of forces is meV/$\rm \AA$. The multi-head attention\cite{vaswani2017attention} mechanism was directly imported from the implementation in PyTorch.}
		\label{table:eps2}
	\end{table}

	
	\subsection{QM9 Dataset}
\begin{figure}[H]
		\centering
		\includegraphics[width=1\linewidth]{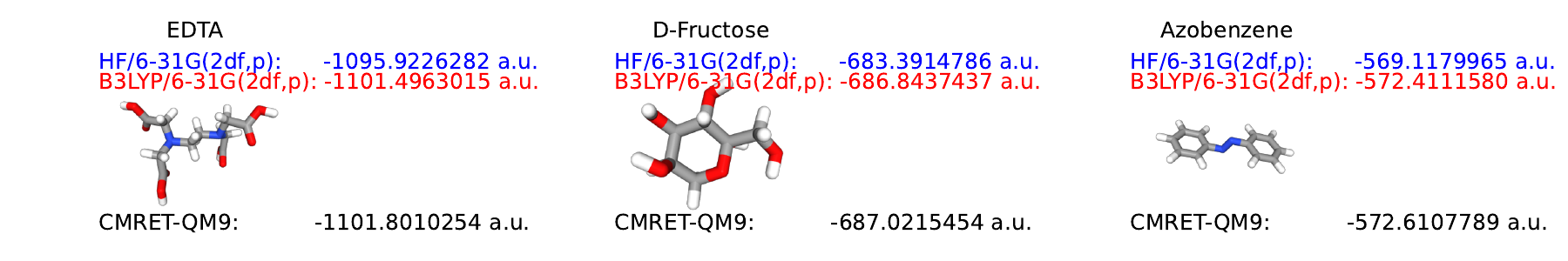}
		\caption{Extrapolated performance compared to HF/6-31G(2df,p) and B3LYP/6-31G(2df,p) total energies.}
		\label{fig:large}
	\end{figure}

	\prettyref{table:qm9} shows the results of our model tested on the QM9\cite{ramakrishnan2014quantum} dataset, in which properties of molecules (up to 9 C, N, O, F `heavy' atoms) were calculated at B3LYP/6-31G(2df,p) level, compared with other models. 110k data were randomly chosen as the training set, while the rest was used for testing (3054 data that failed geometric consistency checks were excluded). Although testing errors on internal energies were in general larger compared to other models, and only comparable to SchNet, QM9-trained CMRET (CMRET-QM9) showed very good extrapolated performance particularly on larger molecules, e.g., EDTA (20 `heavy' atoms), azobenzene (14 `heavy' atoms) and D-fructose (12 `heavy' atoms), compared to HF calculations (\prettyref{fig:large}). This highlights the sensible influence that core, heavier atoms have on the attention element of the transformer when trained on static structure databases like QM9 containing geometry optimised molecules, in agreement with \cite{tholke2022torchmd}.
	 
		\begin{table}[H]
		\centering
		\begin{tabular}{|ll|ccccc|c|}
			\hline
			{} & {unit} & {\small SchNet} & {\small DimeNet++} & {\small PAINN} & {\small Torchmd-NET} & {\small MACE} & {Ours}\\
			\hline
			{$\mu$} & {$D$} & {0.033} & {0.0297} & {0.012} & {0.011} & {0.015} & {\bf 0.0029}\\
			{$\alpha$} & {a$_{0}^{3}$} & {0.235} & {0.0435} & {0.045} & {0.059} & {0.038} & {\bf 0.0137}\\
			{$\epsilon_{\rm HOMO}$} & {meV} & {41} & {24.6} & {27.6} & {\bf 20.3} & {22} & {26.3}\\
			{$\epsilon_{\rm LUMO}$} & {meV} & {34} & {19.5} & {20.4} & {\bf 17.5} & {19} & {20.4}\\
			{$\Delta\epsilon$} & {meV} & {63} & {\bf 32.6} & {45.7} & {36.1} & {42} & {43.9}\\
			{$\langle R^{2}\rangle$} & {a$_{0}^{2}$} & {0.073} & {0.331} & {0.066} & {0.033} & {0.210} & {\bf 0.0278}\\
			{ZPVE} & {meV} & {1.7}  & {\bf 1.21} & {1.28} & {1.84} & {1.23} & {\bf 1.21}\\
			{U$_{0}$} & {meV} & {14} & {6.32} & {5.85} & {6.15} & {\bf 4.1} & {19.02}\\
			{U} & {meV} & {19} & {6.28} & {5.83} & {6.38} & {\bf 4.1} & {18.90}\\
			{H} & {meV} & {14} & {6.53} & {5.98} & {6.16} & {\bf 4.7} & {18.40}\\
			{G} & {meV} & {14} & {7.56} & {7.35} & {7.62} & {\bf 5.5} & {20.34}\\
			{C$_{v}$} & {cal/K/mol} & {0.033} & {0.023} & {0.024} & {0.026} & {0.021} & {\bf 0.0183}\\
			\hline
		\end{tabular}
		\caption{Testing results on QM9 dataset (metric: MAE). Data of SchNet, DimeNet++, PAINN and Torchmd-NET are from P. Th{\"o}lke \textit{et al}\cite{tholke2022torchmd}. Data of MACE were taken from D.P. Kovacs \textit{et al}\cite{Kov_cs_2023}. It took 7 days per task to train our model on single V100 GPU. The best results are in bold. 
		Calculated properties: $\mu$ dipole moment; $\alpha$ isotropic polarizability;  $\epsilon_{\rm HOMO}$, $\epsilon_{\rm LUMO}$ energies of HOMO and LUMO, respectively; $\Delta\epsilon$ HOMO-LUMO gap; $\langle R^{2}\rangle$ electronic spatial extent; ZPVE Zero Point Vibrational Energy; U$_{0}$ internal energy at 0 K; U, H, G, C$_{v}$  internal energy, enthalpy, free energy and heat capacity at 298.15 K, respectively.}
		\label{table:qm9}
	\end{table}


	\subsection{MD17 Dataset}
	The geometries of the molecules in the MD17 dataset were generated via molecular dynamics (MD)\cite{tholke2022torchmd,schutt2021equivariant,doi:10.1126/sciadv.1603015,gasteiger2020directional}. It was suggested by A.S. Christensen \textit{et al}\cite{christensen2020role} that since the geometries are highly correlated, the maximum number of single molecule conformations should be limited to 1,000. Following this advice, we trained our model on 1,000 randomly selected  samples of each molecule. The trained model was tested on 10k samples different from the training ones. The testing results on MD17 dataset compared with SchNet\cite{schutt2017schnet}, PAINN\cite{schutt2021equivariant}, SpookyNet\cite{unke2021spookynet} and Torchmd-NET\cite{tholke2022torchmd} are summarised in \prettyref{table:md17}. High-symmetry molecules like benzene are learned with great accuracy, comparable to the one achived for naphtalene, less symmetric but very similar in terms of chemical bonds. Toluene, with an MAE of 0.093, is also well described, however to a lower precision compared to Torchmd-NET.  While force prediction is improved compared for example to SchNet, it remains slightly inferior to other models on this dataset, in particular for low-symmetry molecules like salicylic acid and its acetylated form, aspirin. 

	\begin{table}[H]
		\centering
		\begin{tabular}{|c|ccccc|c|}
			\hline
			{} & {} & {\small SchNet} & {\small PAINN} & {\small SpookyNet} & {\small Torchmd-NET} & {Ours}\\
			\hline
			\multirow{2}*{\tabincell{c}{\small Benzene\\}} & {\small energy} & {0.08} & {--} & {--} & {0.058} & {\bf 0.045}\\
			{} & {\small force} & {0.31} & {--} & {--} & {0.196} & {\bf 0.075}\\
			\hline
			\multirow{2}*{\tabincell{c}{\small Ethanol\\}} & {\small energy} & {0.08} & {0.064} & {0.052} & {0.052} & {\bf 0.022}\\
			{} & {\small force} & {0.39} & {0.224} & {\bf 0.095} & {0.109} & {0.264}\\
			\hline
			\multirow{2}*{\tabincell{c}{\small Malonaldehyde\\}} & {\small energy} & {0.13} & {0.091} & {0.079} & {0.077} & {\bf 0.074}\\
			{} & {\small force} & {0.66} & {0.319} & {\bf 0.167} & {0.169} & {0.312}\\
			\hline
			\multirow{2}*{\tabincell{c}{\small Naphthalene\\}} & {\small energy} & {0.16} & {0.116} & {0.116} & {0.085} & {\bf 0.083}\\
			{} & {\small force} & {0.58} & {0.077} & {0.089} & {\bf 0.061} & {\bf 0.061}\\
			\hline
			\multirow{2}*{\tabincell{c}{\small Aspirin\\}} & {\small energy} & {0.37} & {0.167} & {0.151} & {0.123} & {\bf 0.115}\\
			{} & {\small force} & {1.35} & {0.338} & {0.258} & {\bf 0.235} & {0.611}\\
			\hline
			\multirow{2}*{\tabincell{c}{\small Salicylic acid\\}} & {\small energy} & {0.20} & {0.116} & {0.114} & {0.093} & {\bf 0.072}\\
			{} & {\small force} & {0.85} & {0.195} & {0.180} & {\bf 0.126} & {0.475}\\
			\hline
			\multirow{2}*{\tabincell{c}{\small Toluene\\}} & {\small energy} & {0.12} & {0.095} & {0.094} & {\bf 0.074} & {0.093}\\
			{} & {\small force} & {0.57} & {0.094} & {0.087} & {\bf 0.067} & {0.254}\\
			\hline
			\multirow{2}*{\tabincell{c}{\small Uracil\\}} & {\small energy} & {0.14} & {0.106} & {0.105} & {0.095} & {\bf 0.090}\\
			{} & {\small force} & {0.56} & {0.139} & {0.119} & {\bf 0.095} & {0.288}\\
			\hline
		\end{tabular}
		\caption{Testing results on MD17 dataset (metric: MAE). Data of SchNet, PAINN and Torchmd-NET are taken from P. Th{\"o}lke \textit{et al}\cite{tholke2022torchmd}, data of SpookyNet from O.T. Unke \textit{et al}\cite{unke2021spookynet}. Models were all trained on 1k data points of each subset of MD17. The training took around 3.5 days on single V100 graphic card to convergence. The best results are in bold. The unit of energy is kcal/mol. The unit of force is kcal/mol/$\rm \AA$.}
		\label{table:md17}
	\end{table}

	
	\subsection{ISO17 Dataset}
	The ISO17 dataset contains MD trajectories of 127 different \ce{C7O2H10} molecules\cite{schutt2017schnet}. This benchmark tests the extrapolation capability, i.e., after training on 80\% of configurations (400k data points) the model is tested on unseen MD trajectories of both seen molecules (10\%) and unseen molecules (10\%). The testing results of our model are shown in \prettyref{table:iso17}. The results for known molecules in unknown conformations largely mirror the observation made above for the MD17 dataset, placing CMRET below SchNet on the MAE scale. However, CMRET can much more reliably extrapolate from known to unknown molecules, a feature that is rapidly deteriorating in SchNet and PhysNet.

	\begin{table}[H]
		\centering
		\begin{tabular}{|c|ccc|c|}
			\hline
			{} & {} & {\small SchNet} & {\small PhysNet} &  {Ours}\\
			\hline
			\multirow{2}*{\tabincell{c}{\small known molecules/\\unknown conformations}} & {\small energy} & {0.36} & {\bf 0.10} & {0.24}\\
			{} & {\small force} & {1.00} & {\bf 0.12} & {0.55}\\
			\hline
			\multirow{2}*{\tabincell{c}{\small unknown molecules/\\unknown conformations}} & {\small energy} & {2.40} & {2.94} & {\bf 1.59}\\
			{} & {\small force} & {2.18} & {1.38} & {\bf 1.15}\\
			\hline
		\end{tabular}
		\caption{Testing results on ISO17 dataset (metric: MAE). SchNet data are taken from K. Sch{\"u}tt \textit{et al}\cite{schutt2017schnet}. Data of PhysNet are sourced from O.T. Unke and M. Meuwly\cite{unke2019physnet}. It took around 29 days (400 epochs) to train the model on single V100 graphic card. The best results are in bold. The unit of energy is kcal/mol. The unit of force is kcal/mol/$\rm \AA$.}
		\label{table:iso17}
	\end{table}
	\subsection{DES370K Dataset}
	
\begin{figure}[ht]
		\centering
		\includegraphics[width=0.9\linewidth]{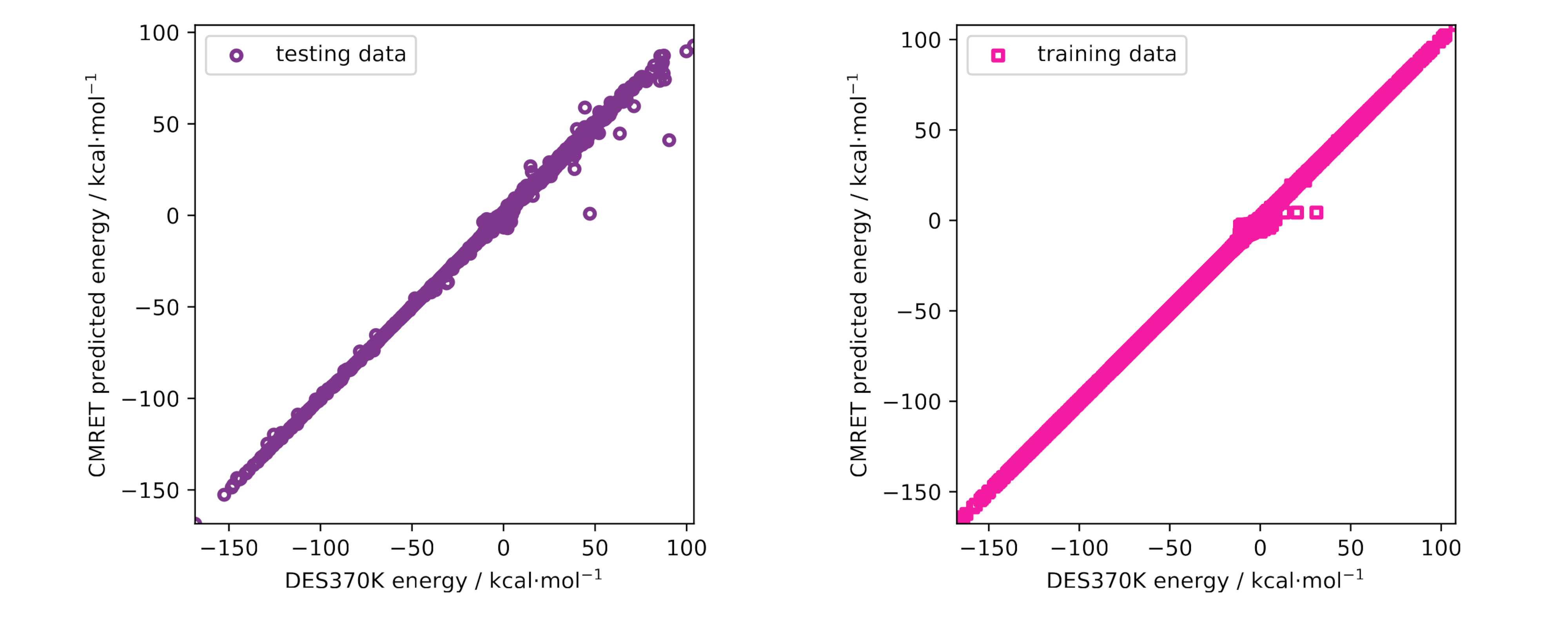}
		\caption{CMRET predicted energies \textit{v.s.} labelled energies on DES370K dataset of testing split (left) and training split (right).}
		\label{fig:des370k}
	\end{figure}

	The DES370K dataset\cite{donchev2021quantum} consists of 370,959 dimer (charged or neutral) geometries labelled with a series of dimer interaction energies (unit in kcal/mol) calculated at CCSD(T)/CBS level of accuracy. A random sample of 80\% of the geometries formed the training set, while the remaining 20\% constituted the testing set. We only trained our model on the total CCSD(T) interaction energies (labelled as `cc\_CCSD(T)\_all' in the dataset). Predicted values \textit{vs.} labelled values are plotted in \prettyref{fig:des370k} (MAE(testing set)=0.137 kcal/mol, MAE(training set)=0.092 kcal/mol). The extrapolative means of CMRET remains reliable over the whole energy range, with only minor deterioration compared to the MAE achieved in the training. Importantly, this test highlights the capability of CMRET to learn intermolecular interactions via the same graph/attention mechanism effective for intramolecular interactions, which adds to its transferability to a broader spectrum of molecules and molecular landscapes, whose total energies result from a mixture of covalent and dispersive interactions, affecting therefore different length ranges. This flexibility is warranted by the attention layer, which can capture long-range interactions outside the cut-off range of functions like \prettyref{eq:cutoff} (see also Section Methods).
	
	\section{Discussion and Conclusions}
	In this work, we introduced a method to embed molecular net charge and spin without adding network parameters, which can be applied to any equivariant neural network that utilizes vector features (e.g., PAINN\cite{schutt2021equivariant}, Torchmd-NET\cite{tholke2022torchmd}, etc.). The purposed model CMRET, employing a modified continuous-filter convolution, achieved higher prediction accuracy on several subset of MD17 and QM9 dataset than recent state-of-the-art models. We showed the importance of selecting proper activation functions and attention temperatures in the self-attention mechanism of the equivariant transformer. We found that a Softmax-based self-attention outperformed a series of ReLU-like functions, deviating from previous works (P. Th{\"o}lke \textit{et al}\cite{tholke2022torchmd}) which suggested that the replacement of Softmax with a ReLU-like function, e.g., Swish function, in the attention architecture would improve the performance. Furthermore, we showed that increasing the attention temperature to $\tau = \sqrt{2d}$ benefited the extrapolation capability for unseen conformations, pinpointing the key role of  attention  in learning non-local, "spooky"~\cite{unke2021spookynet} interactions. This, we argue, allows for an embedding of electronic degrees of freedom already at the initialisation stage. 
	
	CMRET does not introduce any \textit{a priori} angular dependency via specific functions. An angular sensitivity is rather achieved via non-linear mixing of the basis functions in the coordinate frames by means of the attention mechanism. The latter allows to capture long-range interaction outside of the cut-off range that informs the creation of edge features.
	
	The CMRET network is expected to provide a robust starting point for the study of molecular ensembles, with the explicit inclusion of spin states, as well as periodic solids with spin polarisation. As shown in this work, the spin embedding strategy can be easily transferred to other approaches, including its good extrapolation/transferability behaviour, for comparison and further model development.
	
	The superior extrapolation capabilities as well as the flexible learning of intra- and inter-molecular interactions from atomic coordinates, lends CMRET good flexibility for navigating molecular and periodic solid landscapes.  
		
	\subsection{Limitations}
	The extrapolation capability was limited when the training data was highly correlated, e.g., in short MD trajectories. To reduce the computational complexity, we did not implement higher order message passing in this work. The question whether higher order messages benefit this specific equivariant architecture will be studied in a future work.
	
	\section{Acknowledgements}
	Access to computational facilities were provided by Supercomputing Wales. All calculations were performed on (\textit{Hawk}) at Cardiff University.
	We thank the Leverhulme Trust for support under Project No. RPG-2020-052.
	The DFT and HF calculations were preformed via the PySCF\cite{https://doi.org/10.1002/wcms.1340} package.
	Our model CMRET is implemented in PyTorch\cite{NEURIPS2019_9015} . The source code of the model and the data used are open-sourced at \url{https://github.com/Augus1999/torch_CMRET}.
	
	\bibliographystyle{rsc}
	\bibliography{cmret.bib}
	
\end{document}